\documentclass[prc,twocolumn,nofootinbib]{revtex4-2}

\usepackage{color}
\usepackage{graphicx}
\usepackage{dcolumn}
\newcolumntype{d}[1]{D{.}{.}{#1}}
\usepackage{bm}
\usepackage{braket}
\usepackage{amsmath}
\usepackage{here}
\usepackage{color}
\usepackage{array}
\usepackage{enumerate}
\usepackage{amssymb}
\usepackage{bbm}
\usepackage{dsfont}
\usepackage[scr=rsfs]{mathalpha}
\usepackage{longtable}
\def\be{\begin{equation}} 
\def\ee{\end{equation}}

\usepackage{xcolor}
\graphicspath{{figs/}}
\usepackage{ytableau}
\usepackage{braket}
\usepackage[bookmarks=true,colorlinks,%
citecolor=blue,linkcolor=blue,anchorcolor=blue,filecolor=blue,urlcolor=blue,%
           ]{hyperref}

\usepackage[normalem]{ulem}  

\renewcommand\sout{\bgroup\color[rgb]{1,0.75,0.8} \ULdepth=-.5ex \ULset}

\newcommand\T{\rule{0pt}{3ex}}       
\newcommand\B{\rule[-1.5ex]{0pt}{0pt}} 

\begin{document}

\title{
Role of Wigner SU(4) Symmetry in Low-Energy Spin-Isospin Transitions
of Three-Valence-Nucleon Systems on LS Closed Cores}

\author{Takumi Shoji}
\thanks{These authors contributed equally to this work as co-first authors.}
\affiliation{Degree Programs in Pure and Applied Sciences, University of Tsukuba, Tsukuba 305-8571, Japan}

\author{Kotaro Uzawa}
\thanks{These authors contributed equally to this work as co-first authors.}
\affiliation{Nuclear Data Center, Japan Atomic Energy Agency, Tokai 319-1195, Japan}
\affiliation{ 
RIKEN Nishina Center for Accelerator-based Science, RIKEN, Wako 351-0198, Japan
}

\author{Anil Kumar}
\author{Noritaka Shimizu}
\affiliation{Center for Computational Sciences, University of Tsukuba, Tennodai, Tsukuba 305-8577, Japan}

\begin{abstract}
We investigate low-energy Fermi (F) and Gamow--Teller (GT) transition strengths in three-valence-nucleon systems from the viewpoint of Wigner SU(4) symmetry.
For $^{19}{\rm F}$--$^{19}{\rm Ne}$, shell-model calculations show that the ground states contain a large component of the SU(4) representation whose spin-isospin wave function is fully antisymmetric.
By varying the spin--orbit splitting in the shell-model Hamiltonian,
we demonstrate clear correlations of the SU(4) overlap with both the GT strength to the ground state and its fraction of the total GT strength.
For $^{7}{\rm Li}$--$^{7}{\rm Be}$, shell-model calculations show that the $3/2^-_\mathrm{g.s.}$ and $1/2^-_1$ states both belong predominantly to the same supermultiplet as the
$^{19}{\rm F}$--$^{19}{\rm Ne}$ ground states.
The observed strong GT transitions to these states are therefore interpreted as transitions within the same SU(4) multiplet, split by the spin-orbit interaction.
Thus, Wigner SU(4) symmetry provides a unified interpretation of low-energy GT strength concentrations in three-valence-nucleon systems.
In particular, the concentration of both F and GT strengths in the same
$^{19}{\rm F}$--$^{19}{\rm Ne}$ ground-state transition is naturally
understood as a three-valence-nucleon extension of the SU(4) mechanism
underlying the low-energy strong GT peak in
$^{18}{\rm O}$--$^{18}{\rm F}$.
\end{abstract}

\maketitle

\textit{Introduction.}
The nucleon possesses both spin and isospin degrees of freedom, each associated with SU(2) symmetry. 
Fermi (F) and Gamow--Teller (GT) transitions are fundamental examples of spin--isospin transitions \cite{Rapaport1994}.
Beta-decay measurements provide a direct probe of spin--isospin transitions, but are restricted to processes with positive $Q$ values.
In contrast, charge-exchange reactions, such as $(p,n)$ and $(^{3}{\rm He},t)$ can access F and GT transitions over a wide excitation-energy region \cite{Fujita2011}.
Consequently, these reactions have played a central role in experimental studies of giant GT resonances and the quenching of GT transition strengths
\cite{GRtext}.

In contrast to the giant GT resonance, the GT transition strength in some nuclei is concentrated in the low-energy region. 
For example, in the $^{42}{\rm Ca}(^{3}{\rm He},t)^{42}{\rm Sc}$ reaction at $\theta=0^\circ$, the GT transition strength is concentrated in the lowest $1^+$ state at 0.6~MeV \cite{Fujita2014}.
This prominent peak is referred to as the low-energy super-GT (LESGT) state.
Similar low-energy peaks have been observed in the $^{18}{\rm O}(p,n)^{18}{\rm F}$ and $^{18}{\rm O}(^{3}{\rm He},t)^{18}{\rm F}$ reactions \cite{Anderson1983,Mercer1994,Heerden1999,Fujita2019}.
Previous studies have shown that isoscalar $pn$ pairing plays an important role in reproducing both the strengths and energies of these LESGT peaks \cite{Fujita2014,Bai2014,Tanimura2014,Kaneko2018,Uzawa2024}.

It is natural to ask how these low-energy peak structures are extended to three-valence-nucleon systems.
In the $^{19}{\rm F}\to{}^{19}{\rm Ne}$ reactions \cite{rapaport1984,fujita_private}, both F and GT strengths are strongly concentrated in the $J^\pi=1/2^+$ ground-state transition.
In contrast, in $^{7}{\rm Li}\rightarrow{}^{7}{\rm Be}$ the GT strength is shared by $3/2^-_{\text{g.s.}}$ and $1/2^-_{1}$
\cite{Rapaport1990}.

In the present work, our aim is to investigate the characteristic low-energy F and GT strength distributions in  three-valence-nucleon systems.
Unlike the two-valence-nucleon systems, these systems contain an unpaired valence nucleon, and an interpretation based on the pairing correlations is therefore not directly applicable. Therefore, we have performed the  shell-model calculations for the spin--isospin transitions in
the three-valence-nucleon systems. 
We analyze the shell-model wave functions and transition strengths in terms of Wigner SU(4) symmetry to clarify the mechanism underlying the characteristic concentration of GT strength.

\textit{Wigner SU(4) Symmetry.}
Wigner SU(4) symmetry unifies spin SU$_S$(2) and isospin SU$_T$(2) symmetries, with SU$_S$(2)$\times$SU$_T$(2) embedded as a subgroup of SU(4).
Rotations in SU(4) space are generated by the spin operators $\sigma_i \ (i=1,2,3) $, the isospin operators $\tau_j \ (j=1,2,3)$, and the spin-isospin operators $(\sigma_i\tau_j) \ (i,j=1,2,3)$. 
Thus, the total number of generators is 15.
Under SU(4) symmetry, transitions induced by $\boldsymbol{\sigma}$, $\boldsymbol{\tau}$ and $\boldsymbol{\sigma\tau}$ are restricted within each irreducible representation.
The nucleon possesses spin–isospin degrees of freedom,
\begin{equation}
\begin{pmatrix}
\uparrow n \\
\uparrow p \\
\downarrow n \\
\downarrow p
\end{pmatrix},
\end{equation}
and this four-dimensional vector belongs to the fundamental representation of the SU(4) group, denoted by \raisebox{-0.75ex}{$\ydiagram[]{1}$}.
We next consider a three-nucleon system, where each nucleon belongs to the fundamental representation of SU(4).
The irreducible decomposition of the three-particle system is given by \cite{Pfeifer2003}
\begin{equation}
\label{decomposition}
 \raisebox{-0.75ex}{\ydiagram[]{1}}\otimes\raisebox{-0.75ex}{\ydiagram[]{1}}\otimes\raisebox{-0.75ex}{\ydiagram[]{1}} = \raisebox{2.5ex}{\ydiagram[]{1,1,1}}\oplus2\ \ydiagram[]{2,1}\oplus\raisebox{-0.75ex}{\ydiagram[]{3}}.   
\end{equation}
These Young tableaux are labeled by $[f_1,f_2,f_3,f_4]$, where $f_i$ denotes the numbers of boxes in the $i$-th row.
The number of rows is at most four, and therefore four integers are used. 
Their sum, $f_1+f_2+f_3+f_4$, gives the number of fundamental representations in the direct product, and therefore only three of the four integers are independent.
Instead of $[f_1,f_2,f_3,f_4]$, each irreducible representation can also be expressed using three quantum numbers $[P,P',P'']$, defined as \cite{Pang1969, Gaponov1982}
\begin{equation}
\begin{aligned}
    P   &= \tfrac{1}{2}(f_1 + f_2 - f_3 - f_4), \\
    P'  &= \tfrac{1}{2}(f_1 - f_2 + f_3 - f_4), \\
    P'' &= \tfrac{1}{2}(f_1 - f_2 - f_3 + f_4).
\end{aligned}
\end{equation}
The quantities labeling the three irreducible representations in Eq.~(\ref{decomposition}) are summarized in Table~\ref{table1}, along with the spin and isospin quantum numbers $(S,T)$ contained in each representation \cite{Bohr-Mottelson1}.

These numbers are related to the eigenvalue of the SU(4) Casimir operator\footnote{The indices $p$ and $q$ label particles and are not components of the operators.}
\begin{equation}
    C[SU(4)]=\frac{1}{4}\sum_{pq} \left(\boldsymbol{\sigma}_p\cdot\boldsymbol{\sigma}_q+\boldsymbol{\tau}_p\cdot\boldsymbol{\tau}_q+(\boldsymbol{\sigma}_p\boldsymbol{\tau}_p)(\boldsymbol{\sigma}_q\boldsymbol{\tau}_q)\right),
    \label{C2}
\end{equation} as \cite{Gaponov1982}
\begin{equation}
\begin{aligned}
&\langle P, P', P''|C[SU(4)]|P, P', P''\rangle  \\
&\quad =
\left(P^2+4P+P'^2+2P'+P''^2\right).
\end{aligned}
\label{C2_expc}
\end{equation}
This expression is a generalization of the SU(2) spin operator,
\begin{equation}
\langle S|
\frac{1}{4}\sum_{pq}
\boldsymbol{\sigma}_p\cdot\boldsymbol{\sigma}_q
|S\rangle
=
S(S+1).
\end{equation}
Since the SU(4) Casimir term can be interpreted as an SU(4) rotational energy \cite{Gaponov1982}, the ground states are expected to favor the representations with smaller Casimir eigenvalues.
Among the three SU(4) irreducible representations appearing on the right-hand side of Eq.~(\ref{decomposition}), the representation $[f_1,f_2,f_3,f_4]=[1,1,1,0]$, or equivalently $[P,P',P'']=[1/2,1/2,-1/2]$, gives the smallest eigenvalue.

SU(4) symmetry is exact when the coefficients associated with the terms $\boldsymbol{\sigma}\cdot\boldsymbol{\sigma}$, $\boldsymbol{\tau}\cdot\boldsymbol{\tau}$, and $(\boldsymbol{\sigma}\boldsymbol{\tau})\cdot(\boldsymbol{\sigma}\boldsymbol{\tau})$ in the nuclear two-body interaction are identical.
In realistic nuclear many-body Hamiltonians, however, SU(4) symmetry is only approximate. 
It is therefore necessary to quantify the degree to which this symmetry is preserved.

\begin{table*}[htb]
\centering
\begin{tabular}{c|c|c|c}
\hline
\hline
 & $[f_1,f_2,f_3,f_4]$ & $[P,P',P'']$ & $(S,T)$ \\
\hline

\raisebox{0pt}
  [\dimexpr\height+5pt\relax]
  [\dimexpr\depth+5pt\relax]
  {$\vcenter{\hbox{$\ydiagram[]{1,1,1}$}}$}
& $[1,1,1,0]$
& $[1/2,1/2,-1/2]$
& $(1/2,1/2)$ \\
\hline

\raisebox{0pt}
  [\dimexpr\height+5pt\relax]
  [\dimexpr\depth+5pt\relax]
  {$\vcenter{\hbox{$\ydiagram[]{2,1}$}}$}
& $[2,1,0,0]$
& $[3/2,1/2,1/2]$
& $(3/2,1/2),\ (1/2,3/2),\ (1/2,1/2)$ \\
\hline

\raisebox{0pt}
  [\dimexpr\height+5pt\relax]
  [\dimexpr\depth+5pt\relax]
  {$\vcenter{\hbox{$\ydiagram[]{3}$}}$}
& $[3,0,0,0]$
& $[3/2,3/2,3/2]$
& $(3/2,3/2),\ (1/2,1/2)$ \\
\hline
\hline
\end{tabular}
\caption{The partition numbers $f_i$, the three quantum numbers
$[P,P',P'']$, and the spin and isospin quantum numbers $(S,T)$
of the SU(4) multiplets \cite{Bohr-Mottelson1}.}
\label{table1}
\end{table*}

\textit{SU(4) Interpretation of F and GT Transitions in Three-Valence-Nucleon Systems.}
We summarize the experimental information of F and GT strengths of the
$^{19}{\rm F}$--$^{19}{\rm Ne}$,
$^{7}{\rm Li}$--$^{7}{\rm Be}$, and
$^{43}{\rm Ti}$--$^{43}{\rm Sc}$ systems in Table~\ref{tab:mgt}.

\begin{table*}[]
  \centering
  \caption{
    The table lists Fermi transition strengths obtained assuming isospin symmetry and experimental Gamow--Teller transition strengths for $^{19}{\rm F}(1/2^+_{\rm g.s.})\rightarrow^{19}{\rm Ne}$, $^{7}{\rm Li}(3/2^-_{\rm g.s.})\rightarrow^{7}{\rm Be}$, and $^{43}{\rm Ti}(7/2^-_{\rm g.s.})\rightarrow^{43}{\rm Sc}$ transitions. 
    $J_f^\pi$ and $E_{x,f}$ denote the spin-parity and excitation energy of the final state, respectively, and the values are taken from the ENSDF level data \cite{ENSDF}.}
  \label{tab:mgt}

  \setlength{\tabcolsep}{4pt}

  \begin{ruledtabular}
    \begin{tabular}{ccccc@{\qquad}cccc}
      \multicolumn{5}{c}{$^{19}{\rm F}\rightarrow^{19}{\rm Ne}$}
      &
      \multicolumn{4}{c}{$^{7}{\rm Li}\rightarrow^{7}{\rm Be}$}
      \T\B\\
      \cline{1-5}
      \cline{6-9}

      $J_f^\pi$
      & $E_{x,f}$ (MeV)
      & $B({\rm F})$
      & $B({\rm GT})${\cite{rapaport1984}}
      & $B({\rm GT})${\cite{fujita_private}}
      \footnote{
        The values in this column are unpublished data provided by
        Y. Fujita, H. Fujita, and G. Takayama.
      }
      & $J_f^\pi$
      & $E_{x,f}$ (MeV)
      & $B({\rm F})$
      & $B({\rm GT})${\cite{Rapaport1990}}
      \T\B\\

      \cline{1-5}
      \cline{6-9}

      $1/2^+$
      & 0
      & 1
      & 1.64
      & 1.625(5)
      & $3/2^-$
      & 0
      & 1
      & 1.24(1)
      \T\B\\

      $3/2^+$
      & 1.536
      & 0
      & 0.05
      & 0.030(9)
      & $1/2^-$
      & 0.429
      & 0
      & 1.11(5)
      \T\B\\

      $3/2^+$
      & 4.033
      & 0
      & not observed
      & 0.003(1)
      & & & &
      \T\B\\

      $1/2^+$
      & 5.351
      & 0
      & 0.05
      & 0.039(12)
      & & & &
      \T\B\\
    \end{tabular}
  \end{ruledtabular}

  \vspace{1.0em}

  \begin{ruledtabular}
    \begin{tabular}{cccc@{\qquad}cccc}
      \multicolumn{8}{c}{$^{43}{\rm Ti}\rightarrow^{43}{\rm Sc}$}
      \T\B\\
      \cline{1-8}

      $J_f^\pi$
      & $E_{x,f}$ (MeV)
      & $B({\rm F})$
      & $B({\rm GT})${\cite{Honkanen1987}}
      & $J_f^\pi$
      & $E_{x,f}$ (MeV)
      & $B({\rm F})$
      & $B({\rm GT})${\cite{Honkanen1987}}
      \T\B\\

    \cline{1-4}
      \cline{5-8}

      $7/2^-$
      & 0
      & 1
      & 0.663(6)
      & $5/2^-$
      & 2.289
      & 0
      & 0.749(56)
      \T\B\\

      $5/2^-$
      & 0.845
      & 0
      & 0.257(19)
      & $5/2^-$
      & 2.335
      & 0
      & 0.221(17)
      \T\B\\

      $7/2^-$
      & 1.408
      & 0
      & 0.172(13)
      & $(5/2,9/2)^-$
      & 2.459
      & 0
      & 0.371(28)
      \T\B\\

      $(5/2,9/2)^-$
      & 1.883
      & 0
      & 0.120(12)
      & $(5/2\text{--}9/2)^-$
      & 2.760
      & 0
      & 0.209(16)
      \T\B\\

      $(3/2,5/2)^-$
      & 1.963
      & 0
      & 0.044(9)
      & & & &
      \T\B\\
      
    \end{tabular}
  \end{ruledtabular}
\end{table*}
In the $^{19}{\rm F}$--$^{19}{\rm Ne}$ case, the GT transition strength to the ground state, $B({\rm GT};^{19}\rm F(1/2^+_{g.s.})\to{^{19}{\rm Ne}}(1/2^+_{g.s.}))\simeq 1.6$, exhausts a large fraction of the observed low-energy GT strength, whereas transitions to the excited states are strongly suppressed.
The Fermi strength is also concentrated in the $J^\pi=1/2^+$ ground state of  $^{19}{\rm Ne}$, which is the isobaric analogue state (IAS).
In general, the GT operator can induce transitions not only to this ground state but also to excited states  in $^{19}{\rm Ne}$ with $J^\pi=1/2^+$ or $3/2^+$.
The observed concentration of the GT strength in the ground-state transition is therefore a characteristic feature that requires explanation.

In the $^{7}{\rm Li}(p,n)^{7}{\rm Be}$ reaction, strong GT transitions populate both the ground-state IAS and the $1/2^-_1$ state at $E_x=0.429$~MeV.
Unlike the $^{19}{\rm F}$--$^{19}{\rm Ne}$ system, the ground states have $J^\pi=3/2^-$, which may account for the different GT-strength distribution pattern.

Furthermore, in the $^{43}{\rm Ti}$--$^{43}{\rm Sc}$ system, the experimental GT strengths are fragmented over several low-lying states, unlike those in the $^{19}{\rm F}$--$^{19}{\rm Ne}$ and $^{7}{\rm Li}$--$^{7}{\rm Be}$ systems.

The characteristic concentration patterns described above can be understood in terms of SU(4) selection rules.
The $^{19}{\rm F}$ and $^{19}{\rm Ne}$ ground states have $J^\pi=1/2^+$.
If these states are dominated by the \raisebox{2.5ex}{$\ydiagram[]{1,1,1}$} representation with $(S,T)=(1/2,1/2)$ and an $L^\pi=0^+$ orbital component, coupling  $L^\pi=0^+$ with $S=1/2$ gives $J^\pi=1/2^+$.
In the SU(4) limit, the F and GT operators connect only states within the quartet
\begin{equation}
\left\{
 |^{19}{\rm F};1/2^+,J_z=\pm\tfrac{1}{2}\rangle,\,
 |^{19}{\rm Ne};1/2^+,J_z=\pm\tfrac{1}{2}\rangle
\right\},
\end{equation}
which forms a closed subspace for these transitions.

The $3/2^-_{\mathrm{g.s.}}$ and $1/2^-_1$ states of $^{7}$Li and $^{7}$Be involve $p$-shell valence nucleons and are therefore naturally associated with an $L^\pi=1^-$ orbital component.
For the same \raisebox{2.5ex}{$\ydiagram[]{1,1,1}$} representation with $(S,T)=(1/2,1/2)$, coupling $L^\pi=1^-$ with $S=1/2$ produces states with $J^\pi=3/2^-$ and $1/2^-$.
The strong GT transitions to the ground and first excited states can thus be interpreted as transitions within the same SU(4) multiplet, split by the spin-orbit interaction.

\textit{Shell-Model Calculations.}
To investigate the SU(4) symmetry in low-energy F and GT transitions in three-nucleon mirror systems $^{19}\mathrm{F}$--$^{19}\mathrm{Ne}$,  $^{7}\mathrm{Li}$--$^{7}\mathrm{Be}$, and $^{43}\mathrm{Ti}$--$^{43}\mathrm{Sc}$, we have performed the shell model calculations, employing the KSHELL code \cite{SHIMIZU2019372}.

For the $^{19}\mathrm{F}$--$^{19}\mathrm{Ne}$ system, we employ the $sd$-shell model space with the isospin-conserving USDB interaction \cite{Richter2008}.
To quantify SU(4) symmetry, we diagonalize the Casimir operator in Eq.~(\ref{C2}) and project each shell-model eigenstate onto the subspaces belonging to individual SU(4) irreducible representations.
The $^{19}{\rm F}$ and $^{19}{\rm Ne}$ ground states contain a large component of the $[P,P',P'']=[1/2,1/2,-1/2]$ representation, with an overlap probability of 0.881, as shown in Table~\ref{tab:mgt_usdb}.
The two mirror nuclei have identical overlaps because USDB preserves isospin symmetry.
This result is consistent with the good SU(4) symmetry found near closed shells in the $sd$ shell \cite{Kar1983,Vogel1993}.
The values $\langle L^2\rangle/\hbar^2=0.249$ and $\langle S^2\rangle/\hbar^2=0.926$ further indicate dominant $L=0$ and $S=1/2$ components, supporting their coupling to $J^\pi=1/2^+$.

We next discuss the F and GT transition strengths in $^{19}$F. Since the USDB Hamiltonian preserves isospin symmetry, the Fermi strength is exhausted by the transition between the $T=1/2$ ground-state isobaric analogue states, yielding $B(\rm F) = 1$ ,while transitions to excited states vanish.
The bare GT operator in shell model calculations often overestimates the magnitudes of the experimentally extracted GT transition matrix elements, which arise from several effects, including configuration mixing beyond the model space~\cite{SHIMIZU1974282,Bertsch1982,Wakawa1997} and contributions from two-body currents~\cite{Gysbers2019}.

To take into account this effects, an effective GT operator is introduced by multiplying the bare operator by a quenching factor $q$,
\begin{equation}
  (\sigma\tau_{\pm}) \rightarrow q(\sigma\tau_{\pm}),
\end{equation}
with $q=0.77$, consistent with Refs.~\cite{Wildenthal1983,Kumar2020}.
The calculated values of $B({\rm GT})$ are summarized in Table~\ref{tab:mgt_usdb}.
Using the bare GT operator, we calculate the GT strengths summed over all final states in $sd$-shell model space
\begin{align}
 S_- &\equiv
 \sum_f B({\rm GT^-};
 {}^{19}{\rm F}_{\rm g.s.}\rightarrow{}^{19}{\rm Ne}(f))
 =3.37, \label{eq:Sminus}\\
 S_+ &\equiv
 \sum_f B({\rm GT^+};
 {}^{19}{\rm F}_{\rm g.s.}\rightarrow{}^{19}{\rm O}(f))
 =0.37.\label{eq:Splus}
\end{align}
Their difference, $S_--S_+=3.00$, satisfies the Ikeda sum rule.
After applying the quenching factor $q=0.77$, the total strength in the former direction is $q^2S_-=2.00$.
Thus, the ground-state strength, $B({\rm GT})=1.64$ in Table~\ref{tab:mgt_usdb} accounts for $82\%$ of the calculated total strength.
The $3/2^+_1$ and $1/2^+_2$ states of ${}^{19}{\rm Ne}$ also have large overlap probabilities with the fully antisymmetric SU(4) representation.
The $3/2^+_1$ state is dominated by $L=2$, suppressing the GT transition from the predominantly $L=0$ ground state.
The $1/2^+_2$ state, on the other hand, is dominated by $L=0$, similar to the ground state.
As shown in Table~\ref{tab:mgt_usdb}, the $1/2^+_2$ state in
$^{19}\mathrm{Ne}$ has a large overlap with the subspace associated with
the $[P,P',P'']=[1/2,1/2,-1/2]$ representation.
On the other hand, the GT strength for
${}^{19}\mathrm{F}(1/2^+_{\rm g.s.})\rightarrow{}^{19}\mathrm{Ne}(1/2^+_2)$
is small.
In the SU(4) limit with $L=0$,
the $1/2^+_1$ and $1/2^+_2$ states share the same SU(4) spin--isospin structure. 
Therefore, their orbital wave
functions must be orthogonal.
Since the GT operator acts only on the spin--isospin part, 
the GT transition between these states
is forbidden.

\begin{table*}
  \centering
  \caption{
Calculated excitation energies of the final states, SU(4) overlaps, $\langle L^2\rangle/\hbar^2$, $\langle S^2\rangle/\hbar^2$, and F and GT strengths for low-lying states of $^{19}{\rm Ne}$ using the USDB interaction.
The $B({\rm GT})$ values include the quenching factor $q=0.77$.
  }
  \label{tab:mgt_usdb}
  \begin{ruledtabular}
  \begin{tabular}{ccccccc}
    $J^\pi_f$ & $E_{x,f}\,{\rm (MeV)}$
    & $P_{\rm SU(4)}({}^{19}\mathrm{Ne};J^\pi_f)$
    & $\frac{\langle L^2\rangle}{\hbar^2}$
    & $\frac{\langle S^2\rangle}{\hbar^2}$
    & $B({\rm F})$
    & $B({\rm GT})$ \T\B\\
    \hline

    $1/2^+_1$&0
    & 0.881
    & 0.249
    & 0.926
    & 1
    & 1.637\T\B\\

    $3/2^+_{1}$&1.770
    & 0.800
    & 5.904
    & 0.974
    & 0
    & $7\times10^{-5}$ \T\B\\

    $1/2^+_2$ & 6.213
    & 0.807
    & 0.512
    & 0.990
    & 0
    & 0.001 \T\B\\

    $3/2^+_2$ & 6.819
    & 0.315
    & 3.134
    & 1.862
    & 0
    & 0.012 \T\B\\
  \end{tabular}
  \end{ruledtabular}
\end{table*}

To confirm that the concentration of the GT transition strength originates from SU(4) symmetry, we vary the degree of SU(4) symmetry in the USDB Hamiltonian.
In the original USDB Hamiltonian, the single-particle energies of the $d$ orbitals are $\epsilon_{d5/2}=-3.9257~{\rm MeV}$ and $\epsilon_{d3/2}=2.1117~{\rm MeV}$.
We define
\begin{equation}
    \tilde{\epsilon}_d
    =
    \frac{3\epsilon_{d5/2}+2\epsilon_{d3/2}}{5}
    =
    -1.5107 ~{\rm MeV},
\end{equation}
and
\begin{equation}
    \Delta_{\rm LS}
    =
    \epsilon_{d3/2}-\epsilon_{d5/2}
    =
    6.0374~{\rm MeV}.
\end{equation}
We introduce a scaling factor $c$ and vary the spin--orbit splitting as
\begin{equation}
    \epsilon_{d5/2}
    =
    \tilde{\epsilon}_d
    -
    \frac{2}{5}c\Delta_{\rm LS},
\end{equation}
and
\begin{equation}
    \epsilon_{d3/2}
    =
    \tilde{\epsilon}_d
    +
    \frac{3}{5}c\Delta_{\rm LS}.
\end{equation}
Here $c=1$ recovers the original single-particle energies, while increasing $c$ enhances SU(4) symmetry breaking.

\begin{figure}
\includegraphics[width=7cm]{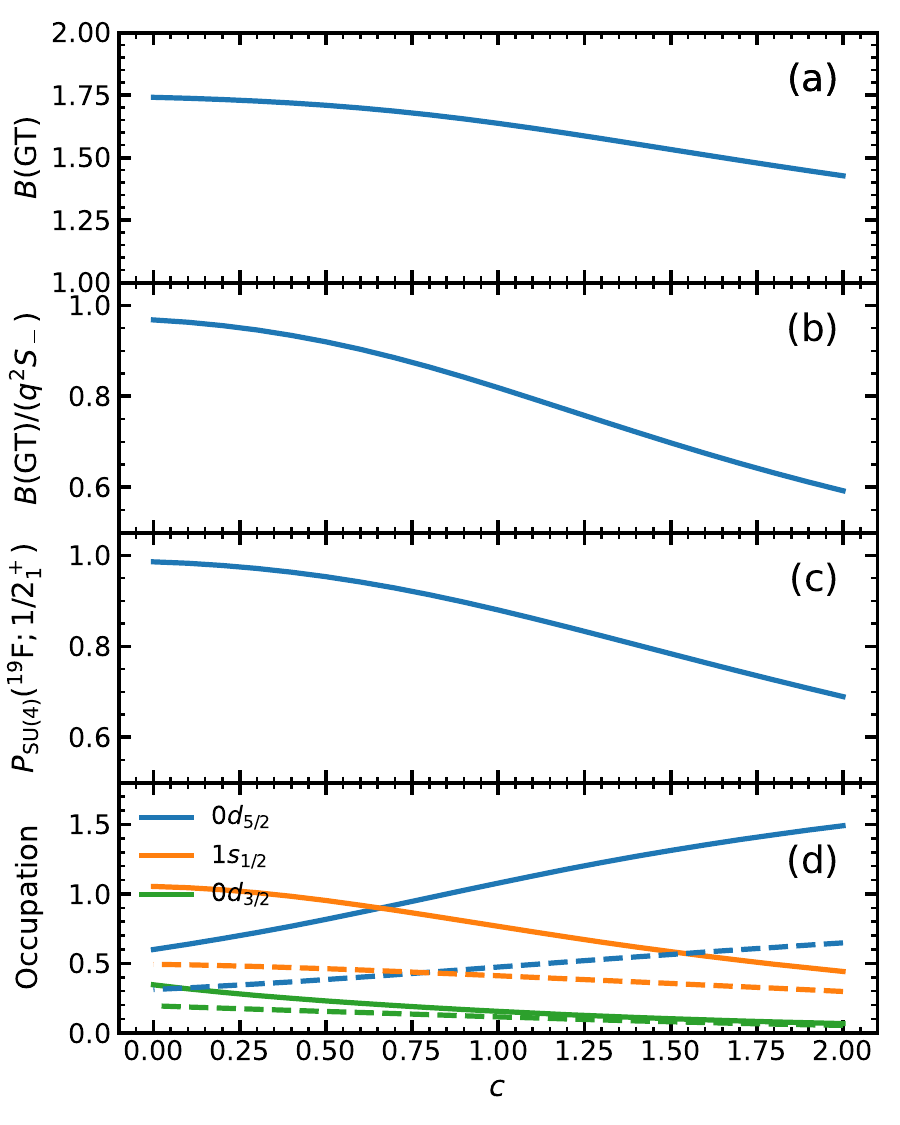}
\caption{
(a) Dependence of
$B(\mathrm{GT};{}^{19}\mathrm{F}(1/2^+_1)
\to{}^{19}\mathrm{Ne}(1/2^+_1))$
on the spin--orbit scaling factor $c$ (see text).
The quenching factor $q=0.77$ is included.
(b) Ratio of the $B(\mathrm{GT})$ value in panel (a) to the
quenched total GT strength, $q^2S_-$.
The quenching factor $q^2$ cancels between the numerator and denominator,
making the ratio independent of $q$.
(c) The SU(4) overlap of the lowest
$1/2^+$ state in $^{19}\mathrm{F}$, as defined in
Eq.~(\ref{overlap}).
(d) Occupation numbers of the $sd$-shell orbitals for the lowest $1/2^+$ state
in $^{19}{\rm F}$ as a function of $c$.
The solid and dashed lines represent neutron and proton occupation numbers,
respectively.}
\label{change_su4}
\end{figure}

Figure~\ref{change_su4}(a) shows
$B(\mathrm{GT};{}^{19}\mathrm{F}(1/2^+_1)
\to{}^{19}\mathrm{Ne}(1/2^+_1))$
as a function of the spin--orbit scaling factor $c$, while panel (b)
shows its ratio to the quenched total GT strength, $q^2S_-$, with
$S_-$ defined in Eq.~(\ref{eq:Sminus}).
At $c=0$, the ground-state transition exhausts nearly all of the
GT strength, whereas increasing the spin--orbit splitting
redistributes a substantial fraction of the strength to excited states.
To quantify the SU(4) symmetry, we define the overlap probability of an arbitrary shell-model eigenstate $|\Psi\rangle$ with an SU(4) irreducible representation $[P,P',P'']$ as
\begin{equation}
 P_{\rm SU(4)}({}^{A}X;J^\pi_n)
 =
 \sum_{i\in[\frac{1}{2},\frac{1}{2},-\frac{1}{2}]}
 \left|
 \left\langle i \middle| \Psi({}^{A}X;J^\pi_n) \right\rangle
 \right|^2 ,
 \label{overlap}
\end{equation}
where $\{|i\rangle\}$ is an orthonormal basis spanning the
Casimir eigenspace associated with
$[P,P',P'']=[1/2,1/2,-1/2]$.
Figure~\ref{change_su4}(c) shows
$P_{\rm SU(4)}({}^{19}\mathrm{F};1/2^+_1)$ as a function of $c$.
The overlap exhibits a dependence on $c$ similar to $B(\mathrm{GT})$ and $B(\mathrm{GT})/(q^2S_-)$.
Their correlated behavior indicates that the loss of ground-state
GT-strength concentration is closely associated with SU(4)-symmetry
breaking.
Finally, Fig.~\ref{change_su4}(d) shows the occupation numbers of the $sd$-shell orbitals.
They vary smoothly with $c$, indicating 
that the correlated variations
shown in Fig.~\ref{change_su4}(a)--(c) do not arise from a drastic
change in the shell configuration.
These results support an SU(4) interpretation of the low-energy GT concentration, analogous to the LESGT states in the two-valence-nucleon systems \cite{Fujita2014}.

For the $^{7}{\rm Li}$--$^{7}{\rm Be}$ system, we employ the $p$-shell model space and the Cohen--Kurath interaction \cite{Cohen1965}.
The results are summarized in Table~\ref{tab:mgt2}.
Although the excitation energy of the $1/2^-_1$ state is higher than the experimental value, the concentration of the $B({\rm GT})$ strength in the two low-lying states is reproduced.
Both states have large overlaps with the same SU(4) representation, $[P,P',P'']=[1/2,1/2,-1/2]$, while $\langle L^2\rangle/\hbar^2\simeq 2$ and $\langle S^2\rangle/\hbar^2\simeq 3/4$ indicate that the $L=1$ and $S=1/2$ components are dominant.
Therefore, the two strong transitions to the $J^\pi=3/2^-$ and $1/2^-$ states can be interpreted as transitions within the same multiplet, split by the coupling between $L=1$ and $S=1/2$ components.

\begin{table}
  \centering
  \caption{
  Similar to Table~\ref{tab:mgt_usdb}, but for the $^{7}{\rm Li}$--$^{7}{\rm Be}$ transitions.
  Here, $P_{\rm SU(4)}$ denotes $P_{\rm SU(4)}({}^{7}\mathrm{Be};J^\pi_f)$.
  The shell-model calculations employ the Cohen--Kurath interaction.
  The calculated $B({\rm GT})$ values include the quenching
  factor $q=0.82$ \cite{Chou1993}.}
  \label{tab:mgt2}
  \begin{ruledtabular}
  \begin{tabular}{ccccccc}
      $J^\pi_f$
    & $E_{x,f}$ (MeV)
    & $P_{\mathrm{SU}(4)}$
    & $\langle L^2\rangle/\hbar^2$
    & $\langle S^2\rangle/\hbar^2$
    & $B({\rm F})$
    & $B({\rm GT})$ \T\B\\
    \hline

    $3/2^-_1$
    & 0
    & 0.955
    & 2.027
    & 0.792
    & 1
    & 1.065 \T\B\\

    $1/2^-_1$
    & 1.979
    & 0.997
    & 2.007
    & 0.756
    & 0
    & 0.865 \T\B\\
  \end{tabular}
  \end{ruledtabular}
\end{table}

\begin{table*}
  \centering
  \caption{
  Similar to Tables~\ref{tab:mgt_usdb} and \ref{tab:mgt2}, but for the
  $^{43}{\rm Ti}$--$^{43}{\rm Sc}$ transitions.
  The shell-model calculations employ the GXPF1A interaction.
  Here, $P_{\rm SU(4)}$ denotes $P_{\rm SU(4)}({}^{43}\mathrm{Sc};J^\pi_f)$.
  The shell-model $B({\rm GT})$ values are multiplied by the square of the quenching factor  $q=0.744$. \cite{Pinedo1996}}
  \label{tab:mgt3}
  \begin{ruledtabular}
  \begin{tabular}{ccccccc}
    $J^\pi_f$
    & $E_{x,f}$ (MeV)
    & $P_{\mathrm{SU}(4)}$
    & $\langle L^2\rangle/\hbar^2$
    & $\langle S^2\rangle/\hbar^2$
    & $B({\rm F})$
    & $B({\rm GT})$ \T\B\\
    \hline

    $7/2^-_1$ & 0
    & 0.546 
    & 12.343
    & 1.4700
    & 1
    & 0.496\T\B\\

    $9/2^-_1$ & 1.771
    & 0.321
    & 18.221
    & 2.020
    & 0
    & 0.180\T\B\\
    
    $5/2^-_1$ &2.280
    & 0.528
    & 10.788
    & 1.182
    & 0
    & 0.540\T\B\\
    
  \end{tabular}
  \end{ruledtabular}
\end{table*}

As a contrasting $pf$-shell example,  we briefly discuss the $^{43}{\rm Ti}$--$^{43}{\rm Sc}$ system.
As shown in Table~\ref{tab:mgt}, the experimental GT strengths for $^{43}{\rm Ti}\rightarrow{}^{43}{\rm Sc}$ are fragmented over several low-lying states, unlike the concentrated patterns observed in the $^{19}{\rm F}$--$^{19}{\rm Ne}$ and $^{7}{\rm Li}$--$^{7}{\rm Be}$ systems.
For $^{43}{\rm Ti}$--${^{43}{\rm Sc}}$ system, we performed shell-model calculations in the $pf$-shell model space using the GXPF1A interaction~\cite{Honma2005}.
The results are given in Table~\ref{tab:mgt3}.
The SU(4) overlaps are substantially smaller than those obtained for the $sd$- and $p$-shell cases, reflecting the strong spin--orbit splitting in the $pf$ shell.
However, the calculated GT strength remains concentrated in a few low-lying states and does not reproduce the observed fragmentation quantitatively.
This discrepancy may indicate missing configuration mixing from core-excited components beyond the present $^{40}{\rm Ca}$ inert-core calculation.
Indeed, previous studies pointed out that low-lying states of $^{43}{\rm Sc}$ cannot be described satisfactorily within a pure $(pf)^3$ space and may involve intruder configurations across the $sd$--$pf$ shell gap \cite{McGrory1973,Poves1981}.
A quantitative calculation including such $^{40}{\rm Ca}$ core excitations is beyond the scope of the present work and is left for future study.

\textit{Summary.}
We investigated low-energy F and GT transition strengths in three-valence-nucleon systems from the viewpoint of Wigner SU(4) symmetry.
Experimentally, the GT strength observed in charge-exchange reactions on $^{19}\mathrm{F}(1/2^+_\mathrm{g.s.}) \to$ $^{19}\mathrm{Ne}$ is strongly concentrated in the $^{19}\mathrm{Ne}(1/2^+_\mathrm{g.s.})$ state.
Shell-model calculations reproduce the suppression of GT transitions to the excited states.
By varying the spin--orbit splitting, we found clear correlations of
the SU(4) overlap with both GT strength to the ground state and its
fraction of the total GT strength, supporting the interpretation that
the observed strength concentration originates from approximately good
SU(4) symmetry.
For the $^{7}{\rm Li}$--$^{7}{\rm Be}$ system, the GT strength is distributed mainly by the $J^\pi=3/2^-_{\rm g.s.}$ and $1/2^-_1$ states.
The shell-model calculation reproduces this concentration in the two low-lying states, both of which have large overlaps with the same SU(4) representation.
The observed strength distribution can therefore be understood as arising from the same SU(4) spin--isospin structure coupled to the $p$-shell orbital angular momentum, which generates the splitting $J^\pi=3/2^-$ and $1/2^-$ states.
We also examined the $^{43}{\rm Ti}$--$^{43}{\rm Sc}$ system as an example in the $pf$-shell.
The calculated SU(4) overlaps are considerably smaller, reflecting the strong spin--orbit splitting in this shell.
Moreover, the calculation in the pure $pf$-shell model space does not reproduce the experimentally observed fragmentation of the GT strength or all of the low-lying levels.
This discrepancy suggests that configurations involving excitations from the inert core may be important.

These results indicate that Wigner SU(4) symmetry provides a useful framework for understanding low-energy spin--isospin strength concentration, especially in light three-valence-nucleon systems where SU(4) symmetry is approximately preserved.
In particular, SU(4) symmetry naturally explains how the concentration of GT strength at low energy in the $^{18}{\rm O}$--$^{18}{\rm F}$ system extends to the ground-state transition carrying both Fermi and GT strengths in the $^{19}{\rm F}$--$^{19}{\rm Ne}$ system.

\begin{acknowledgments}
The authors sincerely thank Y. Fujita, H. Fujita, and G. Takayama for providing unpublished Gamow--Teller transition strengths for the $^{19}{\rm F}$--$^{19}{\rm Ne}$ system.
The authors are also grateful to them for valuable discussions and helpful comments on the experimental data. 
We also acknowledge T. Miyagi for fruitful discussions.
This research was supported by JST SPRING, Japan Grant No. JPMJSP2124, JST ERATO Grant No. JPMJER2304, and KAKENHI (25K07294, 25K00995).
\end{acknowledgments}

\bibliography{ref}

\end{document}